\documentclass[10pt,aps,pr,twocolumn,showpacs,amssymb]{revtex4-1}
\usepackage{amsmath,graphicx,epsfig,mathrsfs}
\usepackage{bm}
\usepackage{epigraph}
\usepackage{geometry}    
\usepackage{ulem}

\usepackage{cases}
\usepackage{cancel}
\usepackage{float}
\newcommand{\qv}{{\bm q}}

\newcommand{\W}{\Omega}

\newcommand{\s}{\sigma}

\newcommand{\R}{{\mathcal R}}

\newcommand{\e}{\varepsilon}

\usepackage{color}

\usepackage{color}
\definecolor{darkgreen}{rgb}{0.0, 0.5, 0.0}
\definecolor{brown}{rgb}{0.59, 0.29, 0.0}
\definecolor{darkgreen2}{rgb}{0.01, 0.75, 0.24}
\usepackage{hyperref}
\hypersetup{backref=true,
pagebackref=true,
hyperindex=true,
breaklinks=true,
urlcolor=cyan,
linkcolor=blue,
bookmarks=true,
colorlinks=true,
citecolor=blue,
bookmarksopen=true}

\begin{document}
\title{Theory of chiral effects in magnetic textures with spin-orbit coupling}
\author{C.  A.  Akosa$^{1,2}$}
\email{collins.akosa@riken.jp}
\author{A. Takeuchi$^3$}
\author{Z. Yuan$^4$}
\author{G. Tatara$^{1,5}$}
\affiliation{$^{1,2}$RIKEN Center for Emergent Matter Science (CEMS), 2-1 Hirosawa, Wako, Saitama 351-0198, Japan}
\affiliation{$^2$Department of Theoretical and Applied Physics, African University of Science and Technology (AUST), Km 10 Airport Road, Galadimawa, Abuja F.C.T, Nigeria}
\affiliation{$^3$Department of Physics and Mathematics, Aoyama Gakuin University, Sagamihara, Kanagawa 252-5258, Japan}
\affiliation{$^4$The Center for Advanced Quantum Studies and Department of Physics, Beijing Normal University, 100875 Beijing, China}
\affiliation{$^5$RIKEN Cluster for Pioneering Research (CPR), 2-1 Hirosawa, Wako, Saitama, 351-0198 Japan}
\date{\today}

\begin{abstract}
We present a theoretical study of two-dimensional spatially and temporally varying magnetic textures in the presence of spin-orbit coupling (SOC) of both the Rashba and Dresselhaus types. We show that the effective gauge field due to these SOCs, contributes to the dissipative and reactive spin torques in exactly the same way as in electromagnetism. Our calculations reveal that Rashba (Dresselhaus) SOC induces a chiral dissipation in interfacial (bulk) inversion asymmetric magnetic materials. Furthermore, we show that in addition to chiral dissipation $\alpha_c$, these SOCs also produce a chiral renormalization of the gyromagnetic ratio $\tilde{\gamma}_c$, and show that the latter is intrinsically linked to the former via a simple relation $\alpha_c = (\tau/\tau_{\rm ex}) \tilde{\gamma}_c$, where $\tau_{\rm ex}$ and $\tau$ are the exchange time and the electron relaxation time, respectively. Finally, we propose a theoretical scheme based on the Scattering theory to calculate and investigate the properties of damping in chiral magnets. Our findings should in principle provide a guide for material engineering of effects related to chiral dynamics in magnetic textures with SOC.
\end{abstract}
\maketitle

\section{Introduction}\label{sec:sec1}

The recent years have witnessed a surge in research in nanoscale magneto-electronics that focuses on the 
utilization of the spin degree of freedom of electrons in combination with its charge, to create new functionalities and devices such as magnetic random access memories, hard drives and sensors \cite{Wolf2001, Zutic2004}. The performance of these devices strongly depends on the dissipation of magnetization dynamics. The latter detects the energy required, the speed and efficiency at which these devices  operate. As a result, the qualitative estimation of damping in magnetic materials is in principle indispensable for piloting and designing alternative materials for different spintronics applications. 

Over the past years, several microscopic theories of magnetization dissipation in which SOC is the mediating interaction via which angular momentum (and energy) is dissipated by the precessing magnetization \cite{Kambersky1976, Gilbert2004, Gilmore2007} have been proposed. Recent theories have highlighted the important role that the \textit{s-d} interaction between the local magnetization and the spins of itinerant electrons play in the dynamics of magnetization \cite{Zhang2004}. Indeed, it has been shown that the interaction between a nonuniform precessing magnetization and spins of itinerant electrons give raise to nonlocal contribution to the Gilbert damping \cite{Zhang2009, Umetsu2012}.

A class of magnetic materials that have attracted enormous research interest owing to their offer of enhanced device performances such as low threshold current density and ultra-fast current-induced domain wall motion \cite{Soumyanarayanan2016} are chiral magnets common in materials with SOC and broken inversion symmetry. It was recently pointed out that magnetization damping in these materials include a \textit{chiral} contribution \cite{Kim2015, Jue2016, Akosa2016a, Freimuth2017, Kim2018}. Even though this prediction is appealing towards the realization of ultra-low damping, little information is known about the relative strength of the chiral with respect to the nonchiral contributions to the damping. Furthermore, the nature of the SOC in chiral magnets determines the type of magnetic texture that can be stabilized in the system. Indeed, it has been shown that an effective chiral energy, i.e. Dzyaloshinskii-Moriya interaction can be derived from a microscopic model of electrons moving in a magnetic texture in the presence of SOC \cite{Kikuchi2016,Koretsune2018}. This chiral energy has been shown to stabilize N\'eel (Bloch) domain walls in systems with Rashba (Dresselhaus) SOC as a result of interfacial (bulk) inversion symmetry breaking. 

In this study, we present a theoretical study of an interplay of Rashba and Dresselhaus SOCs in two-dimensional chiral magnets with spatially and temporally varying magnetization. We propose schemes based on the Green's function formalism and the Scattering theory to qualitatively calculate the chiral damping  and chiral renormalization of the gyromagnetic ratio inherent in these materials. We show that just as in the case for chiral energy, these SOCs induce a chiral damping ($\alpha_c$) and chiral renormalization of the gyromagnetic ratio ($\tilde{\gamma}_c$) that are intrinsically linked via $\alpha_c = (\tau/\tau_{\rm ex}) \tilde{\gamma}_c$, where $\tau_{\rm ex}$ and $\tau$ are the exchange time and the electron relaxation time, respectively. Finally, we elucidate the nature and properties of both the chiral and nonchiral contributions to damping in these materials.

This work is organized as follows. In Sec. \ref{sec:sec2}, we introduce the theoretical model based on the Green's function formalism employed to calculate the spin torque induced by a spatially and temporally varying magnetization in the presence of Rashba SOC and  Dresselhaus SOC. In Sec. \ref{sec:sec3}, we study the corresponding current-induced dynamics in the presence of the torques calculated in the preceding section to obtain analytical expressions and estimates of the chiral damping and chiral renormalization of the gyromagnetic ratio. In Sec. \ref{sec:sec4},  we provide a  scheme based on the Scattering theory to calculate the chiral damping contribution. This scheme is applied in Sec. \ref{sec:sec5} via a tight-binding model to numerically compute and investigate the properties of chiral damping and thus verify our theoretical model.
Finally, in Sec. \ref{sec:sec6}, we provide a summary of the main results in this work.

\section{Theoretical model}\label{sec:sec2}
In this section, we outline the theoretical framework employed to calculate the spin torque induced by a spatially and temporally varying magnetization in the presence of both Rashba SOC and Dresselhaus SOC. The calculated torques are classified into \textit{dissipative} or \textit{reactive} based on whether they are \textit{odd} or \textit{even} under time reversal symmetry.  The dissipative torques contribute to a damping that is proportional to the first order derivative of the magnetization and hence \textit{chiral} by nature \cite{Kim2015, Akosa2016a, Jue2016, Freimuth2017, Kim2018}. The reactive torques contribute to the renormalization of the gyromagnetic ratio which is also \textit{chiral} \cite{Freimuth2017, Kim2018}. Our considerations are based on a two-dimensional inversion asymmetric magnet with spatially and temporarily varying magnetization ${\bf m}({\bm r},t)$ described by the Hamiltonian 
\begin{equation}\label{eq:ham0}
\hat{\mathcal{H}} = \frac{\hat{\bf p}^2}{2m} + J {\bf m}({\bm r}, t)\cdot\hat{\boldsymbol{\sigma}} + \hat{\mathcal{H}}_{\rm so},               
\end{equation}
where $m$ is the mass of electron, $\hat{\bf p}$ is the momentum operator, $J$ is the \textit{s-d} exchange coupling between the 
local moment ${\bf m}$ and the electrons with spin represented by the vector of Pauli matrices $\hat{\boldsymbol{\sigma}}$. 
The third term on the right hand side of Eq.~(\ref{eq:ham0}) represents an interplay of Rashba SOC due to interfacial inversion symmetry breaking \cite{Bychkov1984} and Dresselhaus SOC due to bulk inversion symmetry breaking \cite{Dresselhaus1955} given as
\begin{subequations}
\begin{eqnarray}
\hat{\mathcal{H}}_{\rm R} &=& \frac{\beta_{\rm R}}{\hbar} \big( \hat{\sigma}_y \hat{p}_x - \hat{\sigma}_x \hat{p}_y\big)
\end{eqnarray}
and
\begin{eqnarray}
\hat{\mathcal{H}}_{\rm D} &=& \frac{\beta_{\rm D}}{\hbar} \big( \hat{\sigma}_x \hat{p}_x - \hat{\sigma}_y \hat{p}_y \big), 
\end{eqnarray}
\end{subequations}
of strength $\beta_{\rm R}$ and $\beta_{\rm D}$, respectively. In the case of magnetic textures, the exchange term in Eq.~(\ref{eq:ham0}) includes off-diagonal terms. 
This term is diagonalized via a local unitary transformation in the spin space $\hat{U}({\bm x},t) ={\bf n}({\bm x},t)\cdot\hat{\boldsymbol{\sigma}}$, i.e., $\hat{U}^\dagger ({\bf m}\cdot {\bm \sigma}) \hat{U} = \sigma^z$, where ${\bf n} = (\cos\phi \sin\frac{\theta}{2}, \sin\phi \sin\frac{\theta}{2}, \cos\frac{\theta}{2})$\cite{Tatara1994a,Tatara1994b}. In the transformed space (rotating frame with the spin quantization axis along ${\bf m}({\bm r}, t)$), the electrons sees a background of a uniform ferromagnetic state that is \textit{coupled} to the corresponding spin gauge fields due to (i) the texture ${\bf A}_s^\mu$ and (ii) the SOC ${\bf A}_{\rm so}^\mu$, given as
\begin{subequations}
\begin{equation}
{\bf A}_s^\mu  = -\frac{i \hbar}{2e} \mbox{Tr}\Big[\hat{\sigma}^\mu \hat{U}^\dagger {\bm \nabla} \hat{U}\Big]
= \frac{\hbar}{e} ({\bf m} \times {\bm \nabla} {\bf m})^\mu 
\end{equation}
and
\begin{equation}
{\bf A}_{\rm so}^\mu = \frac{m}{e \hbar} {\bm \lambda}_{\rm so}^\nu \R^{\mu \nu} ,
\end{equation}
\end{subequations}
respectively, where ${\bm \lambda}_{\rm so}^\mu$ is given as
\begin{equation}
\begin{pmatrix}
\lambda_{{\rm so},x}^x & \lambda_{{\rm so},x}^y & \lambda_{{\rm so},x}^z\\
\lambda_{{\rm so},y}^x & \lambda_{{\rm so},y}^y  & \lambda_{{\rm so},y}^z \\
\lambda_{{\rm so},z}^x & \lambda_{{\rm so},z}^y & \lambda_{{\rm so},z}^z
\end{pmatrix}
=
\begin{pmatrix}
\beta_{\rm D} & \beta_{\rm R}  & 0 \\
-\beta_{\rm R} & -\beta_{\rm D} & 0 \\
0 & 0 & 0
\end{pmatrix}
\end{equation}
and $\mathcal{R}^{\mu\nu}$ are components of the rotation matrix given by 
\begin{equation}\label{eq:rot}
\mathcal{R}^{\mu\nu}  = 2 n^\mu n^\nu -\delta^{\mu\nu}.
\end{equation}
Furthermore, this unitary transformation modifies spin-dependent observables such as the spin torque and the nonequilibrium spin density of itinerant electrons. In particular, 
the nonequilibrium spin density in the transformed ($\tilde{\bm s}$) and original (${\bm s}$)  frames transforms as
\begin{equation}
s^\mu = \mathcal{R}^{\mu \nu}\tilde{s}^\nu.
\end{equation}
We recall that the presence of non-equilibrium spin density ${\bm s}$ regardless of its source in a magnetic system, exerts a torque ${\bf T}$ on the local magnetization ${\bf m}$ given as 
\begin{equation}\label{eq:sd}
{\bf T} = \frac{Ja_0^2}{\hbar}{\bf m}\times {\bm s},
\end{equation}
where $a_0$ is the lattice constant, $\hbar$ is the reduced Planck's constant. Therefore, to calculate the spin torque on the local magnetization, it suffices to calculate $\tilde{\bm s}$. In this study, we focus on the time-varying magnetization as the primary source of $\tilde{\bm s}$. We treat the interaction between the spin gauge fields ${\bf A}^\mu = {\bf A}_{\rm so}^\mu + {\bf A}_{\rm s}^\mu$  and the background conduction electrons in the transformed frame to be weak, this allows us to apply the perturbation theory to calculate $\tilde{\bm s}$. In particular, we consider the adiabatic limit in which the spins of electrons follow the direction of the local magnetization, and calculate $\tilde{\bm s}$ via the Green's function approach \cite{Tatara2018}, in which the spin gauge fields ${\bf A}^\mu$ are treated perturbatively (see Appendix \ref{sec:appena} for details). Since this work focuses on chiral effects, we consider only up to first order in the spin gauge fields due to SOC. The  relevant contributions to the spin torque induced by the time-dependent texture is calculated using Eq.~(\ref{eq:sd}) as (see Appendix \ref{sec:appena} for details)
\begin{eqnarray}\nonumber \label{eq:torque} 
{\bf T} &=&  \mathcal{C}_1 \big(\partial_t{\bf A}_{\rm so}^{||}\cdot \boldsymbol{\nabla}\big){\bf m} + \mathcal{C}_2{\bf m}\times\big(\partial_t{\bf A}_{\rm so}^{||}\cdot \boldsymbol{\nabla}\big){\bf m}    \\ \nonumber
&+& \mathcal{C}_3  \Big[\big( {\bf m}\times\partial_t{\bf m}\big)\cdot \big( {\bf A}_{\rm so}^{\perp, \mu} \cdot \boldsymbol{\nabla}\big) {\bf m} \Big]{\bm e}_\mu \\
&+&  \mathcal{C}_4 \Big[\big( {\bf m}\times\partial_t{\bf m}\big) \cdot \big(   {\bm \lambda}^\mu_{\rm so} \cdot \boldsymbol{\nabla}\big) {\bf m}\Big]{\bm e}_\mu  + {\bf T}_{\rm nl},
\end{eqnarray}
where the \textit{in-plane} and \textit{out-of-plane} components of the SOC-induced spin gauge fields are given as 
\begin{equation}
{\bf A}_{\rm so}^{||} = {\bm \lambda}_{\rm so}^\mu m^\mu \hspace{2 mm} \mbox{and} \hspace{2 mm} {\bf A}_{\rm so}^{\perp, \mu} = \epsilon^{\mu \nu c}m^\nu {\bm \lambda}_{\rm so}^c.
\end{equation}
The last term on the right hand side of Eq.~(\ref{eq:torque}) represents other contributions to the torque given as 
\begin{equation}\label{eq:nl}
{\bf T}_{\rm nl} =  \big(\mathcal{C}_5 +   \mathcal{C}_6 {\bf m}\times \big) \big({\bf A}_{\rm so}^{||} \cdot \boldsymbol{\nabla} \big)\partial_t {\bf m}. 
\end{equation}
In domain walls, even though ${\bf T}_{\rm nl}$ is locally finite, it vanishes upon the integration over space.
The torque pre-factors in Eqs.~(\ref{eq:torque}) and (\ref{eq:nl}), are given as 
\begin{subequations}\label{eq:cis}
\begin{eqnarray}
\mathcal{C}_1 &=& -  \frac{1}{4\pi }\frac{ma_0^2}{\hbar^2}  \frac{\varepsilon_{\rm F} J^2 (J^2 - \eta^2)}{\eta (J^2 +\eta^2)^2} \\
\mathcal{C}_2 &=&  \frac{1}{2\pi }\frac{ma_0^2}{\hbar^2} \frac{\varepsilon_{\rm F} J^3}{ (J^2 +\eta^2)^2} \\
\mathcal{C}_3 &=& -  \frac{1}{4\pi }\frac{ma_0^2}{\hbar^2}  \frac{\varepsilon_{\rm F} J^2 (J^2 + 3\eta^2)}{\eta (J^2 +\eta^2)^2} \\
\mathcal{C}_4 &=& - \frac{1}{2\pi }\frac{ma_0^2}{\hbar^2}  \frac{\varepsilon_{\rm F} J \eta^2}{ (J^2 +\eta^2)^2}, \\
\mathcal{C}_5 &=& -\frac{1}{\pi }\frac{ma_0^2}{\hbar^2}  \frac{\varepsilon_{\rm F} J^2 \eta }{ (J^2 +\eta^2)^2}, \\
\mathcal{C}_6  &=& -\frac{1}{2\pi }\frac{ma_0^2}{\hbar^2}  \frac{\varepsilon_{\rm F} J (J^2 - \eta^2)}{ (J^2 +\eta^2)^2},
\end{eqnarray}
\end{subequations}
where $\eta = \hbar/2\tau$, $\tau$ being the elastic relaxation time of conduction electrons.  Notice from Eq.~(\ref{eq:cis}) that $\mathcal{C}_{1}$ and $\mathcal{C}_3 \gg \mathcal{C}_{2}$ and $\mathcal{C}_{4}$ and thus, dissipative torque effects are dominant over the reactive torque effects in chiral domain walls.

Observe that Eq.~(\ref{eq:torque}) includes torque terms that are both dissipative ($\propto$ $\mathcal{C}_1$ and $\mathcal{C}_3$) and reactive ($\propto$ $\mathcal{C}_2$ and $\mathcal{C}_4$) based on their symmetry under time reversal. Interestingly,  Eq.~(\ref{eq:torque}) which constitutes one of the main result of this study, shows that in the presence of relaxation \cite{Tatara2013}, the first two terms of the torque takes the same form of the adiabatic  ($\propto \mathcal{C}_1$) and the nonadiabatic ($\propto \mathcal{C}_2$) spin transfer torque proportional to $({\bf E}\cdot {\bm \nabla}){\bf m}$ and ${\bf m}\times({\bf E}\cdot {\bm \nabla}){\bf m}$, respectively \cite{Zhang2004, Thiaville2005}, where ${\bf E}$ is the applied electric field expressed in terms of the electromagnetic vector potential ${\bf A}$ (i.e. ${\bf E} = - \partial_t{\bf A}$). In fact, our result indicates that the effective gauge field of \textit{any} origin contributes to the torque in exactly the same way as the electromagnetic gauge field. Even though this is as expected from symmetry point of view, what is significant is that the \textit{spin transfer} torque arising from the gauge field due to spin-orbit interaction indeed has a nature of a damping torque, as the gauge field is linear in magnetization.

\section{Current-induced chiral magnetization dynamics}\label{sec:sec3}
The previous section was devoted to establishing the nature of the spin torque that itinerant electrons exert on the local magnetization as a result of a time-dependent background magnetization in the presence of SOC. In this section, we provide analytic expressions and a qualitative estimates of the chiral contribution to both the damping and the gyromagnetic ratio. To achieve this, we investigate the influence on dynamics of chiral domain walls via the incorporation of Eq.~(\ref{eq:torque}) into the equation of motion of the magnetization described by the extended Landau-Lifshiftz-Gilbert (LLG) equation 
\begin{equation}\label{eq:llg}
\partial_t{\bf m} = -\gamma {\bf m}\times{\bf H}_{\rm eff}  + \alpha_0{\bf m}\times\partial_t{\bf m} + {\bf T},
\end{equation}
where for completeness we have included the phenomenological Gilbert damping with constant $\alpha_0$,
$\gamma$ is the gyromagnetic ratio, ${\bf H}_{\rm eff}= -\frac{1}{\mu_0M_s}\frac{\partial \mathcal{E}}{\partial {\bf m}} $ is the effective field, $\mathcal{E}$ is the energy density, $M_s$ the saturation magnetization and $\mu_0$ the permeability of free space. We consider a one-dimensional Walker domain wall with  magnetization parametrized by the domain wall centre $X_{\rm c}$ and tilt angle $\phi$, and given in spherical coordinate as ${\bf m} = (\cos\phi\sin\theta, \sin\phi\sin\theta, \cos\theta)$, where $\theta(x) = 2\tan^{-1}\big(\exp\big(s\frac{x-X_{\rm c}}{\lambda_{\rm dw}}\big)\big)$, $s = +1(-1)$ for $\uparrow \downarrow (\downarrow\uparrow)$ 
domain wall,  $\phi = \phi(t)$ and $\lambda_{\rm dw}$ is the width of the wall. The dynamics of the wall is given by coupled equations 
 \begin{subequations}\label{eq:llg}
 \begin{equation}
\big(1  + s\bar{\gamma}_c \big) \partial_t \phi  +  \alpha_0 \frac{s\partial_t{\rm X}_c}{\lambda_{\rm dw}} = - \Gamma_\theta
\end{equation}
and
\begin{equation}
 \big( \alpha_0  +  s\alpha_c \big) \partial_t \phi -  \frac{ s\partial_t{\rm X}_c}{\lambda_{\rm dw}}  = \Gamma_\phi,
\end{equation}
 \end{subequations}
 where 
\begin{equation}
\Gamma_{\theta(\phi)} = \frac{1}{2}\gamma \int_{-\infty}^{+\infty}{\bf H}_{\rm eff}\cdot{\bf e}_{\theta(\phi)} {\rm d} x ,
\end{equation}
${\bf e}_\theta = (\cos\phi \cos\theta, \sin\phi \cos\theta, -\sin\theta)$ and ${\bf e}_\phi = (-\sin\phi, \cos\phi , 0)$. The terms $\alpha_c$ and $\bar{\gamma}_c$ in Eq.~(\ref{eq:llg}) represent the chiral damping  and chiral renormalization of the gyromagnetic ratio and given as 
\begin{subequations}
\begin{equation}\label{eq:dam}
\alpha_c = \frac{ \pi n_{\rm F}  \beta_{\rm so}}{4\hbar \lambda_{\rm dw}} \frac{\tau}{ 1 + \tau^2_{\rm ex}/\tau^2}\cos ( \phi + \phi_{\rm so})
\end{equation}
and 
\begin{equation}\label{eq:gam}
\bar{\gamma}_c = \frac{\pi n_{\rm F} \beta_{\rm so}}{4\hbar \lambda_{\rm dw}}  \frac{\tau_{\rm ex} }{ 1 + \tau^2_{\rm ex}/\tau^2}\cos ( \phi + \phi_{\rm so}) ,
\end{equation}
\end{subequations}
respectively, where $ n_{\rm F} = \nu a_0^2 \varepsilon_{\rm F}$ is the number of conduction electrons at the Fermi level, 
\begin{subequations} \label{eq:soc}
\begin{equation}\label{eq:soca}
\beta_{\rm so} = \sqrt{\beta_{\rm R}^2 + \beta_{\rm D}^2}
\end{equation}
 and 
 \begin{equation}\label{eq:socb}
  \phi_{\rm so} = \tan^{-1}(\beta_{\rm D}/\beta_{\rm R})
 \end{equation}
 \end{subequations}
characterizes the strength of the effective SOC present in the material. 

Eqs.~(\ref{eq:dam}) and (\ref{eq:gam}) constitute one of the main result of this work, from which we infer that: (i) Chiral damping and chiral renormalization of the gyromagnetic ratio are Fermi-surface effects since they are proportional to the the number of available conduction electrons at the Fermi level. (ii) The chiral damping constant is proportional to elastic relaxation time of electrons (i.e. $\alpha_c \propto \tau$) that is well described by the SOC mediated \textit{breathing Fermi surface} mechanism for magnetization relaxation \cite{Kambersky1970, Korenman1974, Kunes2002}. It is worthy to note here that the source of electron relaxation can be from scattering with impurity or domain wall itself and hence $\tau$ should in principle depend on the domain wall width $\lambda_{\rm dw}$ and therefore makes the dependence of the $\alpha_c$ on the domain wall width a bit subtle. (iii) The chiral renormalization of the gyromagnetic ratio is inversely proportional to exchange strength (i.e. $\tilde{\gamma}_c \propto  1/J$)  since $\tau_{\rm ex} = \hbar/2J$ and therefore, is more significant in weak ferromagnets. (iv) The chiral damping and gyromagnetic ratio renormalization as related via 
\begin{equation}\label{eq:rat}
\alpha_c = (\tau/\tau_{\rm ex}) \tilde{\gamma}_c.
\end{equation}
This simple relation provides a means by which one effect can be deduced with the knowledge of the other. 
It turns out that similar correspondence has been established by Kim \textit{et. al.} \cite{Kim2015b}, in the context of texture-induced intrinsic nonadiabaticity in the absence of SOC. For a realistic estimate of these effects, we consider typical material parameters such as $\beta_{\rm so} = 2 \times 10^{-11}$ eV m, $\tau = 1\times 10^{-14}$ s, $\tau_{\rm ex} = 1\times 10^{-15}$ s, $\lambda_{\rm dw} = 10 $ nm and $n_{\rm F} = 1$, from which we obtain  $\alpha_c  = 3\times 10^{-2}$ and $\gamma_c = 3\times 10^{-3}$. In general, for real ferromagnetic materials, $\tau_{\rm ex}/\tau \ll 1$, therefore from Eq.~(\ref{eq:rat}), it is expected that in chiral magnets, chiral damping constitute the dominant mechanism that detects the dynamics of chiral domain walls \cite{Akosa2016a, Jue2016}. Now that we have established the analytical form of the dissipative torque given by Eq.~(\ref{eq:torque}), and the corresponding estimate of the chiral damping and chiral gyromagnetic ratio given by Eqs.~(\ref{eq:dam}) and (\ref{eq:gam}), respectively, in what follows, we use the well established Scattering theory of magnetization dissipation based on the conservation of energy \cite{Brataas2008, Brataas2011} to compliment our analytical calculations and propose a scheme to numerically compute the damping in chiral magnets.

\section{Magnetization Damping from the Scattering theory}\label{sec:sec4}
In what follows, we compliment our analytical treatment of the preceding sections by providing a scheme based on the Scattering theory of magnetization damping to calculate the nonchiral and chiral damping (and hence the chiral renormalization of the gyromagnetic ratio by virtue of Eq. (\ref{eq:rat})). We focus on dissipative torque terms in Eq. (\ref{eq:torque}) and neglect the chiral renormalization of the gyromagnetic ratio (i.e. torque terms that are even under time reversal symmetry). 
However,  notice that effects associated with the chiral renormalization of the gyromagnetic ratio can be straightforwardly inferred from our calculations via the Eq.~(\ref{eq:rat}) which establishes a simple relation between chiral damping and chiral gyromagnetic ratio renormalization due to SOC.
The dynamics of magnetization is well described by the extended Landau-Lifshiftz-Gilbert equation given by 
\begin{equation}\label{eq:llg_gd}
\partial_t{\bf m} = -\gamma {\bf m}\times{\bf H}_{\rm eff}  + \alpha_0{\bf m}\times\partial_t{\bf m} + {\bf T}_{\rm dp},
\end{equation}
${\bf T}_{\rm dp}$ is the dissipative contribution to the torque given in Eq.~(\ref{eq:torque}). Again, we consider a one-dimensional Walker domain wall parametrized by the domain wall centre $X_{\rm c} = X_c(t)$ and tilt angle $\phi =\phi(t)$. Furthermore, since the Scattering theory of magnetization dissipation is based on the conservation of energy, we first calculate the rate of change of the magnetic energy density from Eq.~(\ref{eq:llg_gd}) as
\begin{equation}\label{eq:er0}
\frac{{\rm d} \mathcal{E}}{{\rm d} t} = - \frac{\mu_0M_s}{\gamma} \Big( \alpha_0\partial_t{\bf m} +  {\bf T}_{\rm dp}\times {\bf m}\Big)\cdot \partial_t{\bf m},
\end{equation}
where the negative sign shows that energy is lost by the magnetic system.  Notice that the right hand side of Eq.~(\ref{eq:er0}) is bilinear in $\partial_t{\bf m}$ and can therefore can be re-written in the form 
\begin{equation}\label{eq:ed}
\frac{{\rm d} \mathcal{E}}{{\rm d} t} \equiv  \mathcal{D}_{\rm o} \big(\partial_t\phi\big)^2 + \mathcal{D}_{\rm m} \partial_t\phi \partial_tX_c + \mathcal{D}_{\rm i} \big(\partial_tX_c\big)^2, 
\end{equation}
where $\mathcal{D}_{\rm o}$, $\mathcal{D}_{\rm i}$ and $\mathcal{D}_{\rm m}$ represents the \textit{out-of-plane}, \textit{in-plane} and \textit{mix} dissipation, respectively.  
The substitution of Eq.~(\ref{eq:torque}) into Eq.~(\ref{eq:er0}) yields 
\begin{subequations}
\begin{eqnarray}\label{eq:dcc}
\mathcal{D}_{\rm o} &=& \frac{\mu_0 M_s}{\gamma} \Big( \alpha_0  + s \alpha_c \sin\theta\Big) \sin^2\theta \\ \label{eq:mix}
\mathcal{D}_{\rm m} &=& -\frac{\mu_0 M_s}{\gamma } \tilde{\alpha}_c \cos\theta \sin^3\theta \\ 
\mathcal{D}_{\rm i} &=& \frac{\mu_0 M_s}{\gamma} \frac{\alpha_0}{\lambda_{\rm dw}^2}\sin^2\theta 
\end{eqnarray}
\end{subequations}
where $\alpha_c$ is the chiral damping defined in Eq.~(\ref{eq:dam}) and $\tilde{\alpha}_c$ represent a $\frac{\pi}{2}$-phase shift in $\phi$ of $\alpha_c$ (i.e. $\tilde{\alpha}_c(\phi) = \alpha_c(\phi - \pi/2)$)

Interestingly, SOC induces in addition to the in-plane and out-of-plane damping,  a \textit{mix} term $\mathcal{D}_{\rm m}$ which is \textit{locally} finite as shown in Eq.~(\ref{eq:mix}).  Even though in principle, the spatial integration of $\mathcal{D}_{\rm m}$ vanishes, nonequilibrium dynamics of the magnetization might result to a finite value and hence renormalizes the overall contribution of the chiral damping. However, such corrections are expected to be small and hence, we neglect this effect in the rest of this study. The total rate of energy loss by the magnetic system with cross sectional area $\mathcal{A}$ is given as 
\begin{eqnarray}\label{eq:ener}
\frac{{\rm d} {E}}{{\rm d} t} &=& \mathcal{A} \int_{-\infty}^{+\infty}  \frac{{\rm d} {\mathcal{E}}}{{\rm d} t}   {\rm d} x.
\end{eqnarray}
Following the representation of Eq.~(\ref{eq:ed}), Eq. (\ref{eq:ener}) can be re-written in the form 
\begin{equation}\label{eq:dd}
\frac{{\rm d} {E}}{{\rm d} t} = D_{\rm o} \big(\partial_t\phi\big)^2  + D_{\rm i} \big(\partial_t X_c\big)^2, 
\end{equation}
where
\begin{equation}
D_{{\rm o}({\rm i})} = \mathcal{A} \int_{-\infty}^{+\infty}  \mathcal{D}_{{\rm o}({\rm i})}   {\rm d} x
\end{equation}
and after performing the integration, we obtain 
\begin{subequations}\label{eq:st1}
\begin{eqnarray}
D_{\rm o} &=& \frac{2 \mu_0 M_s \mathcal{A} \lambda_{\rm dw} }{\gamma} \Big(\alpha_0   + s\alpha_c  \Big)
\end{eqnarray}
and
\begin{eqnarray}
D_{\rm i} &=& \frac{2 \mu_0 M_s \mathcal{A} \lambda_{\rm dw}}{\gamma\lambda_{\rm dw}^2} \alpha_0 .
\end{eqnarray}
\end{subequations}
The application of the scattering theory of magnetization dissipation in which, the magnetic system is considered to be at a constant temperature, and the energy loss by the magnetic system is equal
to the total energy pumped into the system  yields \cite{Brataas2008, Foros2008, Hals2009, Brataas2011, Yuan2014}
\begin{equation}\label{eq:ec}
\frac{{\rm d} {E}}{{\rm d} t} = \frac{\hbar}{4\pi }\mbox{Tr}\Big(\frac{{\rm d} \mathcal{S}}{{\rm d} t} \frac{{\rm d} \mathcal{S}^\dagger}{{\rm d} t} \Big), 
\end{equation}
where $\mathcal{S}$ is the scattering matrix at the Fermi energy. Furthermore, since $\mathcal{S} = \mathcal{S}({\bf m})$, we have that $\mathcal{S} = \mathcal{S}(X_{\rm c}(t), \phi(t))$ and therefore Eq.~({\ref{eq:ec}) is transformed into
\begin{equation}\label{eq:ee}
\frac{{\rm d} E}{{\rm d} t}  = \mathcal{A}_{\rm o} \big(\partial_t\phi\big)^2  + \mathcal{A}_{\rm i} \big(\partial_t X_c\big)^2
\end{equation}
where 
\begin{subequations}
\begin{equation}
\mathcal{A}_{\rm o} = \frac{\hbar}{4}\mbox{Tr} \Big(\frac{\partial \mathcal{S}}{\partial \phi} \frac{\partial \mathcal{S}^\dagger}{\partial \phi} \Big)
\end{equation}
and 
\begin{equation}
\mathcal{A}_{\rm i} = \frac{\hbar}{4}\mbox{Tr} \Big(\frac{\partial \mathcal{S}}{\partial X_c} \frac{\partial \mathcal{S}^\dagger}{\partial X_c} \Big)
\end{equation}
\end{subequations}
are proportional to the out-of-plane and in-plane contribution to damping, respectively. Next, comparing Eq.~(\ref{eq:dd}) and Eq.~(\ref{eq:ee}), we have that 
\begin{subequations}\label{eq:st}
\begin{equation}
D_{\rm o} = \mathcal{A}_{\rm o} = \frac{\hbar}{4}\mbox{Tr} \Big(\frac{\partial \mathcal{S}}{\partial \phi} \frac{\partial \mathcal{S}^\dagger}{\partial \phi} \Big) 
\end{equation}
and
\begin{equation}
D_{\rm i} =  \mathcal{A}_{\rm i} = \frac{\hbar}{4}\mbox{Tr} \Big(\frac{\partial \mathcal{S}}{\partial X_{\rm c}} \frac{\partial \mathcal{S}^\dagger}{\partial X_{\rm c}} \Big).
\end{equation}
\end{subequations}
Finally, we obtain the expression of the out-of-plane damping using Eq.~(\ref{eq:st1}) and Eq.~(\ref{eq:st}) as
\begin{equation}\label{eq:cd}
\alpha_0   + s\alpha_c  = C \mbox{Tr}\Big(\frac{\partial \mathcal{S}}{\partial \phi} \frac{\partial \mathcal{S}^\dagger}{\partial \phi} \Big),
\end{equation}
where 
\begin{equation}
C  = \frac{\gamma \hbar }{8 \mu_0 M_s \mathcal{A} \lambda_{\rm dw}}.
\end{equation}
Eq.~(\ref{eq:cd}) provides a very transparent way to extract both the \textit{nonchiral} and \textit{chiral} contribution of the damping. Indeed, since $s = \pm 1$ for $\uparrow\downarrow (\downarrow\uparrow)$ domain walls, the non-chiral and chiral contribution of damping can be computed as
\begin{subequations}
\begin{equation}\label{eq:nch}
\alpha_0   = \frac{C}{2} \mbox{Tr}\left(\frac{\partial \mathcal{S}_{\uparrow\downarrow}}{\partial \phi} \frac{\partial \mathcal{S}_{\uparrow\downarrow}^\dagger}{\partial \phi} + \frac{\partial \mathcal{S}_{\downarrow\uparrow}}{\partial \phi} \frac{\partial \mathcal{S}_{\downarrow\uparrow}^\dagger}{\partial \phi}\right) 
\end{equation}
and
\begin{equation}\label{eq:ch}
\alpha_c   = \frac{C}{2} \mbox{Tr}\left(\frac{\partial \mathcal{S}_{\uparrow\downarrow}}{\partial \phi} \frac{\partial \mathcal{S}_{\uparrow\downarrow}^\dagger}{\partial \phi} - \frac{\partial \mathcal{S}_{\downarrow\uparrow}}{\partial \phi} \frac{\partial \mathcal{S}_{\downarrow\uparrow}^\dagger}{\partial \phi}\right), 
\end{equation}
\end{subequations}
respectively. Therefore, the calculation of nonchiral, chiral and by extension chiral renormalization of the gyromagnetic ratio requires the knowledge of the derivative of the scattering matrix with respect to the domain wall tilt angle $\phi$. The derivation of a close form analytic expressions of the scattering matrix in the presence of SOC is non-trivial even though asymptotic expressions have been derived in the limits  $k_{\rm F}\lambda_{\rm dw} \gg 1$ \cite{Tatara2000} and $k_{\rm F}\lambda_{\rm dw} \ll 1$ \cite{Dugaev2003, Dugaev2005, Duine2009}, where $k_{\rm F}$ is the Fermi wave number.
Therefore, in the following section, we calculate these damping contributions by numerically computing the derivatives of the scattering matrix and its conjugate with respect to the tilt angle $\phi$ of a domain wall to ascertain the correctness of the theoretical treatment presented above.

\begin{figure}[t!]
\centerline{\includegraphics[width=8.2cm]{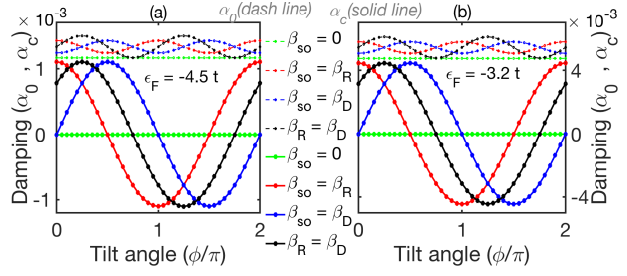}}
\caption{Shows the $\phi$-dependence of the \textit{nonchiral} (dash lines) and \textit{chiral} (solid lines) contribution to the damping in the presence of different SOC. Results shows that $\alpha_c$ is SOC-driven and proportional to the Fermi energy evident in the smaller amplitude for (a) $\varepsilon_{\rm F} = -4.5 t$ compare to (b) $\varepsilon_{\rm F} = -3.2t$. In all calculations with SOC, $\beta_{\rm so} = 0.02 t$.}\label{fig:fig1}
\end{figure}
\section{Numerical Results}\label{sec:sec5}
In this section, we follow the procedure outline in the preceding section and numerically compute the non-chiral and chiral contributions to the damping. To achieve this, we consider a two-dimensional tight-binding model on a square lattice with lattice constant $a_0$. In our calculations, we consider a scattering region of size $1001\times101a_0^2$ to ensure that a domain wall of with $\lambda_{\rm dw} = 15 a_0$ fully relaxed into a ferromagnet at the contact with the left and right leads. The scattering matrix and its derivatives are calculated with the help of KWANT package \cite{Groth2014} from which the nonchiral and chiral contributions of the damping are extracted based on Eqs. (\ref{eq:nch}) and  (\ref{eq:ch}), respectively. Furthermore, in all our calculations, we consider an exchange constant of $J = -2t/3$ and an onsite energy $\varepsilon_i = 0$. The damping parameters are calculated based on the material parameters $M_s = 8\times 10^5\mbox{ A}\mbox{m}^{-1}$, $a_0 = 0.35 \mbox{ nm} $.

\begin{figure}[t!]
\centerline{\includegraphics[width=8.2cm]{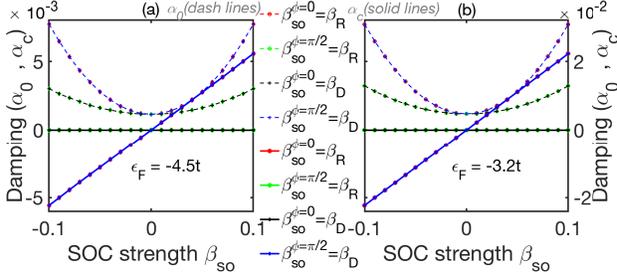}}
\caption{Dependence of chiral and nonchiral damping on the strength of the SOC for (a) $\varepsilon_{\rm F} = -4.5 t$ and  (b) $\varepsilon_{\rm F} = -3.2t$. Notice that the blue and red curves as well as the green and black curves are superimposed showing that the Dresselhaus SOC influences the damping in Bloch walls ($\phi = \pi/2$) exactly the same way that Rashba SOC influences it in N\'eel walls ($\phi = 0$).}\label{fig:fig2}
\end{figure}

Our numerical results of the $\phi$-dependence of the \textit{nonchiral} (dash lines) and \textit{chiral} (solid lines) contributions to the damping in the presence of different SOC for different transport energies: $\varepsilon_{\rm F} = -4.5t$ in Fig. \ref{fig:fig1}\textcolor{blue}{(a)} and  $\varepsilon_{\rm F} = -3.2t$ in Fig. \ref{fig:fig1}\textcolor{blue}{(b)} are in good agreement with our analytical predictions given by Eq.~(\ref{eq:dam}). Indeed, the relative increase in the strength of $\alpha_c$ in Fig. \ref{fig:fig1}\textcolor{blue}{(b)} compared to Fig. \ref{fig:fig1}\textcolor{blue}{(a)} shows that the effect is a Fermi energy effect i.e. $\propto \varepsilon_{\rm F}$. Furthermore, in the absence of SOC, i.e $\beta_{\rm so} = 0$ (green curves), $\alpha_c = 0$ and $\alpha_0$ is  a constant. In the presence of SOC, we considered three interesting cases namely:
(i) $\beta_{\rm so} = \beta_{\rm R}$, i.e., $\beta_{\rm D} = 0$ (red curves) and from Eq.~(\ref{eq:socb}), $\phi_{\rm so} = 0$, therefore $\alpha_c \propto \cos\phi$. 
(ii) $\beta_{\rm so} = \beta_{\rm D}$, i.e., $\beta_{\rm R} = 0$ (blue curves) similarly, $\phi_{\rm so} = \pi/2$, therefore $\alpha_c \propto \sin\phi$. 
(iii) $\beta_{\rm R} = \beta_{\rm D}$ (black curves) and using similar arguments, $\phi_{\rm so} = \pi/4$, therefore $\alpha_c \propto \cos(\phi + \pi/4)$. It is worth mentioning here that in the presence of SOC, the nonchiral damping $\alpha_0$ shows a small oscillatory $\propto \cos^2(\phi + \phi_{\rm so})$ as a result of small SOC-induced anisotropic magnetoresistance. The complete description of the $\phi$-dependence of $\alpha_c$ presented here should in principle provide a guide for material engineering of effects related to damping in chiral magnets. The validity of our analytical model is strengthen with the result of the investigation of the dependence of chiral damping on the strength of the SOC. Indeed, Figs. \ref{fig:fig2} \textcolor{blue}{(a)} and \textcolor{blue}{(b)} show that (i) the non-chiral damping $\alpha \propto \beta_{\rm so}^2$ \cite{Kambersky1970} (ii) the chiral damping $\alpha_c \propto \beta_{\rm so}$ and (iii) chiral and non-chiral damping are Fermi energy effects i.e. $\propto \varepsilon_{\rm F}$. This again, in agreement with our analytical prediction of Eq.~(\ref{eq:dam}). Observe that the Dresselhaus SOC which stabilizes Bloch walls in materials with bulk inversion symmetry breaking affects chiral damping in these materials in exactly the same way that the Rashba SOC that favors N\'eel in materials with interfacial inversion symmetry breaking interaction affects the chiral damping in in these materials. Furthermore, Dresselhaus (Rashba) SOC induces no chiral contribution N\'eel (Bloch) as a result of the $\sin\phi  (\cos\phi)$ symmetry of the chiral damping. Therefore our work shows that the symmetry of the SOC-induced chiral damping is inherited from the symmetry of the materials.

\section{Conclusions}\label{sec:sec6}
We have carried out a detailed theoretical investigation of nature of spin torque and the corresponding dynamics generated by a two-dimensional spatially and temporally varying chiral magnetic textures in the presence of both Dresselhaus and Rashba SOCs. We employed the Green's function formalism to derive expressions for the nonequilibrium spin density and hence the spin torque generated by a spatially and temporally varying chiral magnetic textures in which the gauge field induced by these SOCs is treated perturbatively. Our result indicates that the effective gauge field associated with these SOCs, and by extension of any origin, contributes to the torque in exactly the same way as the electromagnetic gauge field. 
In order to investigate the impact these torques have on the dynamics of chiral magnets, we then incorporated the calculated torques into the LLG equation that governs the dynamics of the magnetization and derived analytic expressions for both the chiral damping $\alpha_c$ and the chiral renormalization of the gyromagnetic ratio $\tilde{\gamma_c}$ and show that $\alpha_c = (\tau/\tau_{\rm ex})\tilde{\gamma}_c$, where $\tau_{\rm ex}$ and $\tau$ are the exchange and electron relaxation times, respectively. Furthermore, we propose a theoretical scheme based on the scattering matrix formalism to calculate and investigate the properties of damping in chiral magnets. Our findings should in principle provide a guide for material engineering of effects related to damping in chiral magnets.

G.T. acknowledges financial support from Grant-in-Aid for Exploratory Research (No.16K13853), Grant-in-Aid for Scientific Research (B) (No. 17H02929) from Japan Society for the Promotion of Science (JSPS), and the Graduate School Materials Science in Mainz (DFG GSC 266). A. T. acknowledges financial support from Grant-in-Aid for Scientific Research (No. 17H02924) from JSPS. Z. Y. acknowledges financial support from the National Natural Science Foundation of China (Grants No. 61774018 and No. 11734004), the Recruitment Program of Global Youth Experts, and the Fundamental Research Funds for the Central Universities (Grant No. 2018EYT03). 
C. A. A. thanks A. Abbout and Y. Yamane for useful discussions.

\twocolumngrid

\href{}{}

\allowdisplaybreaks[4]

\onecolumngrid

\appendix
\section{Non-equilibrium spin density calculation}\label{sec:appena}
In this section, we present a detailed calculation of the non-equilibrium spin density induced by a time-varying magnetization.
To calculate the non-equilibrium spin density $\tilde{\bm s}$, we treat the spin gauge fields ${\bf A}^\mu = {\bf A}_{\rm s}^\mu +{\bf A}_{\rm so}^\mu$ perturbatively
in the adiabatic limit of slow dynamics  ($\hbar \Omega \ll \varepsilon_{\rm F}$) and smooth variation of the magnetization ($q \ll k_{\rm F}$), where $\Omega$, $q$  and  $k_{\rm F}$ are the frequency, the wavenumber, and the Fermi wavenumber, respectively. To simplify notation and render our analysis trackable, we define the Green's functions
\begin{subequations}
\begin{eqnarray}
g_{{\bm k}, \omega} &=& \frac{1}{2}\sum_{\sigma = \pm} \big( 1 + \sigma \sigma^z\big)g_{{\bm k}, \omega, \sigma}, \\ 
g^{r}_{{\bm k}, \omega, \sigma} &=& \frac{1}{\hbar\omega  - \varepsilon_{\bm k} + \varepsilon_{\rm F} + \sigma J + i\eta }, 
\end{eqnarray}
\end{subequations}
such that $g^{r}_{{\bm k}, \omega, \sigma} = (g^{a}_{{\bf k}, \omega, \sigma})^*$ and $\eta = \hbar / 2 \tau$, where $\tau$ is the elastic relaxation time of conduction electrons. The non-equilibrium spin density is defined up to linear order in $\W$ as
\begin{eqnarray}
\tilde{s}^\mu({\bm q},t)
&=&
\frac{e \hbar^2}{2 \pi m}
\sum_{{\bm k}, {\bm q}', {\bm q}''} \partial_t A_i^\nu({\bm q}',t) {\rm Tr} \Big[
k_i \hat{\sigma}^\mu g^{\rm r}({\bm k}+\tfrac{\bm q}{2}, {\bm k}+\tfrac{{\bm q}''}{2})
\hat{\sigma}^\nu g^{\rm a}({\bm k}+\tfrac{{\bm q}'' -{\bm q}'}{2}, {\bm k}-\tfrac{\bm q}{2})
\Big]
\nonumber
\\
&&
+\frac{e^2 \hbar}{2 \pi m}
\sum_{{\bm k}, {\bm q}'} \partial_t \Big[ A_{{\rm s},i}^\nu({\bm q}',t) A_{{\rm so},i}^\nu({\bm q} -{\bm q}',t) \Big]
{\rm Tr} \Big( \hat{\sigma}^\mu g^{\rm r}_{\bm k} g^{\rm a}_{\bm k} \Big),
\end{eqnarray}
where $g^{\rm r (a)}({\bm k}, {\bm k}')$ is the retarded (advanced) Green's function represented by
$g^{\rm r (a)}({\bm k}, {\bm k}) \equiv g^{\rm r (a)}_{\bm k} = (1/2) \sum_{\sigma = \pm} (1 + \sigma \sigma^z) g^{\rm r (a)}_{{\bm k}, \sigma}$,
with $g^{\rm r}_{{\bm k}, \sigma} = (g^{\rm a}_{{\bm k}, \sigma})^* = 1/(-\varepsilon_{\bm k} +\varepsilon_{\rm F} -\sigma J +i \eta)$.
The dominant contributions are linear in $q$ and $\lambda$, and they are depicted in Fig. \ref{fig:app}.
\begin{figure}[t!]
\centerline{\includegraphics[width=8.2cm]{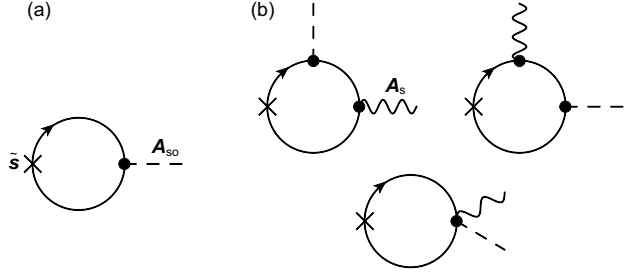}}
\caption{Diagrammatic representation of the non-equilibrium spin density $\tilde{\bm s}$.
The solid, wavy, and dashed lines represent the Green's function, spin gauge potential ${\bf A}_{\rm s}$ and gauge potential due to SOC ${\bf A}_{\rm so}$, respectively. (a) First order and (b) second order in ${\bf A}$ contributions to the non-equilibrium spin density. }\label{fig:app}
\end{figure}
\subsection{First order in ${\bf A}$}
Up to first order in ${\bf A}$, the diagrams that contributes to the non-equilibrium  spin density is given by Fig. \ref{fig:app}\textcolor{blue}{(a)}, from which the components of the spin density are computed as  
\begin{eqnarray}
\tilde{s}^\mu({\bm q},t)
&=&
\frac{e \hbar^2}{2 \pi m}
\partial_t A_{{\rm so}, i}^\nu(\qv,t)
\sum_{\bm k} {\rm Tr} \Big(
k_i \hat{\sigma}^\mu g^{\rm r}_{{\bm k}+\frac{\bm q}{2}}
\hat{\sigma}^\nu g^{\rm a}_{{\bm k}-\frac{\bm q}{2}}
\Big)
\nonumber
\\
&\simeq&
-\frac{i e \hbar^4}{2 \pi m^2}
q_j \partial_t A_{{\rm so}, i}^\nu(\qv,t)
\sum_{\sigma = \pm} \sum_{\bm k} k_i k_j \Big\{
\delta^{\mu z} \delta^{\nu z}
{\rm Im} \Big[ g^{\rm r}_{{\bm k}, \sigma} (g^{\rm a}_{{\bm k}, \sigma})^2 \Big]
\nonumber
\\
&&
+\Big[ \delta^{\mu x} \Big( \delta^{\nu x} {\rm Im} +\sigma \delta^{\nu y} {\rm Re} \Big)
+\delta^{\mu y} \Big( \delta^{\nu y} {\rm Im} -\sigma \delta^{\nu x} {\rm Re} \Big) \Big]
g^{\rm r}_{{\bm k}, -\sigma} (g^{\rm a}_{{\bm k}, \sigma})^2
\Big\}.
\end{eqnarray}

\subsection{Second order in ${\bf A}$}
For completeness we also calculated the second order in ${\bf A}$ contribution to the non-equilibrium spin density as depicted in Fig. \ref{fig:app} \textcolor{blue}{(b)} as 
\begin{eqnarray}
\tilde{s}^\mu({\bm q},t)
&=&
\frac{e^2 \hbar^3}{2 \pi m^2}
\sum_{{\bm q}'}
\Big[ \partial_t A_{{\rm s},i}^\nu({\bm q}',t) A_{{\rm so},j}^o({\bm q}-{\bm q}',t)
+\partial_t A_{{\rm so},i}^\nu({\bm q}',t) A_{{\rm s},j}^o({\bm q}-{\bm q}',t) \Big]
\nonumber
\\
&&
\times \sum_{\bm k} {\rm Tr} \Big(
k_i k_j \hat{\sigma}^\mu g^{\rm r}_{\bm k} \hat{\sigma}^\nu g^{\rm a}_{\bm k} \hat{\sigma}^o g^{\rm a}_{\bm k}
+k_i k_j \hat{\sigma}^\mu g^{\rm r}_{\bm k} \hat{\sigma}^o g^{\rm r}_{\bm k} \hat{\sigma}^\nu g^{\rm a}_{\bm k}
\Big)
\nonumber
\\
&&
+\frac{e^2 \hbar}{2 \pi m}
\sum_{{\bm q}'} \partial_t \Big[ A_{{\rm s},i}^\nu({\bm q}',t) A_{{\rm so},j}^\nu({\bm q}-{\bm q}',t) \Big]
\sum_{\bm k} {\rm Tr} \Big( \hat{\sigma}^\mu g^{\rm r}_{\bm k} g^{\rm a}_{\bm k} \Big)
\nonumber
\\
&=&
\frac{e^2 \hbar^3}{\pi m^2}
\sum_{{\bm q}'}
\Big[ \partial_t A_{{\rm s},i}^\nu({\bm q}',t) A_{{\rm so},j}^o({\bm q}-{\bm q}',t)
+\partial_t A_{{\rm so},i}^\nu({\bm q}',t) A_{{\rm s},j}^o({\bm q}-{\bm q}',t) \Big]
\sum_{\sigma = \pm} \sigma \sum_{\bm k} k_i k_j \bigg(
\nonumber
\\
&&
-\delta^{\mu z} \Big\{
(\delta^{\nu o} -\delta^{\nu z} \delta^{o z})
{\rm Re} \Big[ g^{\rm r}_{{\bm k},\sigma} (g^{\rm a}_{{\bm k},\sigma})^2
-g^{\rm r}_{{\bm k},\sigma} g^{\rm a}_{{\bm k},-\sigma} g^{\rm a}_{{\bm k},\sigma} \Big]
-\sigma \epsilon^{\nu o z}
{\rm Im} \Big( g^{\rm r}_{{\bm k},\sigma} g^{\rm a}_{{\bm k},-\sigma} g^{\rm a}_{{\bm k},\sigma} \Big)
\Big\}
\nonumber
\\
&&
+\delta^{\mu x} \Big[
\delta^{\nu z} \Big( \delta^{o x} {\rm Re} +\sigma \delta^{o y} {\rm Im} \Big)
g^{\rm r}_{{\bm k},\sigma} g^{\rm a}_{{\bm k},-\sigma} g^{\rm a}_{{\bm k},\sigma}
+\delta^{o z} \Big( \delta^{\nu x} {\rm Re} -\sigma \delta^{\nu y} {\rm Im} \Big)
g^{\rm r}_{{\bm k},-\sigma} (g^{\rm a}_{{\bm k},\sigma})^2
\Big]
\nonumber
\\
&&
+\delta^{\mu y} \Big[
\delta^{\nu z} \Big( \delta^{o y} {\rm Re} -\sigma \delta^{o x} {\rm Im} \Big)
g^{\rm r}_{{\bm k},\sigma} g^{\rm a}_{{\bm k},-\sigma} g^{\rm a}_{{\bm k},\sigma}
+\delta^{o z} \Big( \delta^{\nu y} {\rm Re} +\sigma \delta^{\nu x} {\rm Im} \Big)
g^{\rm r}_{{\bm k},-\sigma} (g^{\rm a}_{{\bm k},\sigma})^2
\Big]
\bigg).
\end{eqnarray}
The dominant contributions of the $x$ and $y$ components of the non-equilibrium spin density, $\tilde{s}^x$ and $\tilde{s}^y$, are obtained as
\begin{eqnarray}
\tilde{s}^x
&=&
-\frac{e \hbar^2}{2 \pi m} \partial_i \partial_t A_{{\rm so},i}^\nu
\sum_{\sigma=\pm} \Big( \delta^{\nu x} {\rm Im} +\sigma \delta^{\nu y} {\rm Re} \Big) C_{1,\sigma}
\nonumber
\\
&&
+\frac{e^2 \hbar}{\pi m}
\Big( \partial_t A_{{\rm s},i}^\nu A_{{\rm so},i}^o +\partial_t A_{{\rm so},i}^\nu A_{{\rm s},i}^o \Big)
\nonumber
\\
&&
\times \sum_{\sigma = \pm} \sigma \Big[
\delta^{\nu z} \Big( \delta^{o x} {\rm Re} +\sigma \delta^{o y} {\rm Im} \Big) C_{2,\sigma}
+\delta^{o z} \Big( \delta^{\nu x} {\rm Re} -\sigma \delta^{\nu y} {\rm Im} \Big) C_{1,\sigma}
\Big],
\\
\tilde{s}^y
&=&
-\frac{e \hbar^2}{2 \pi m} \partial_i \partial_t A_{{\rm so},i}^\nu
\sum_{\sigma=\pm} \Big( \delta^{\nu y} {\rm Im} -\sigma \delta^{\nu x} {\rm Re} \Big) C_{1,\sigma}
\nonumber
\\
&&
+\frac{e^2 \hbar}{\pi m}
\Big( \partial_t A_{{\rm s},i}^\nu A_{{\rm so},i}^o +\partial_t A_{{\rm so},i}^\nu A_{{\rm s},i}^o \Big)
\nonumber
\\
&&
\times \sum_{\sigma = \pm} \sigma \Big[
\delta^{\nu z} \Big( \delta^{o y} {\rm Re} -\sigma \delta^{o x} {\rm Im} \Big) C_{2,\sigma}
+\delta^{o z} \Big( \delta^{\nu y} {\rm Re} +\sigma \delta^{\nu x} {\rm Im} \Big) C_{1,\sigma}
\Big],
\end{eqnarray}
where $C_{1(2), \sigma}$ are calculated as
\begin{eqnarray}
C_{1,\sigma}
&=&
\sum_{\bm k} \varepsilon_{\bm k} g^{\rm r}_{{\bm k},-\sigma} (g^{\rm a}_{{\bm k},\sigma})^2
\nonumber
\\
&\simeq&
-\nu \int_{-\infty}^\infty{d\varepsilon}
\frac{\varepsilon}{(\varepsilon -\varepsilon_{\rm F} -\sigma J -i \eta) (\varepsilon -\varepsilon_{\rm F} +\sigma J +i \eta)^2}
\nonumber
\\
&=&
-\frac{i \pi \nu}{2}
\frac{\varepsilon_{\rm F} +\sigma J +i \eta}{(\sigma J +i \eta)^2}
\nonumber
\\
&=&
-\frac{\pi \nu}{2 (J^2 +\eta^2)^2}
\Big[ \eta (J^2 +\eta^2 +2 \sigma \varepsilon_{\rm F} J) +i \varepsilon_{\rm F} (J^2 -\eta^2) +i \sigma J (J^2 +\eta^2) \Big],
\\
C_{2,\sigma}
&=&
\sum_{\bm k} \varepsilon_{\bm k} g^{\rm r}_{{\bm k},\sigma} g^{\rm a}_{{\bm k},-\sigma} g^{\rm a}_{{\bm k},\sigma}
\nonumber
\\
&\simeq&
-\nu \int_{-\infty}^\infty{d\e}
\frac{\varepsilon}{(\varepsilon -\varepsilon_{\rm F} +\sigma J -i \eta) (\varepsilon -\varepsilon_{\rm F} -\s J +i \eta) (\varepsilon -\varepsilon_{\rm F} +\sigma J +i \eta)}
\nonumber
\\
&=&
\frac{\pi \nu}{2 \eta}
\frac{\varepsilon_{\rm F} -\sigma J +i \eta}{\sigma J -i \eta}
\nonumber
\\
&=&
-\frac{\pi \nu}{2 \eta (J^2 +\eta^2)}
\Big( J^2 +\eta^2 -\sigma \varepsilon_{\rm F} J -i \varepsilon_{\rm F} \eta \Big).
\end{eqnarray}
The effective magnetic field ${\bf H}_{\rm eff}^*$ due to this nonequilibrium spin density is given by 
\begin{eqnarray}
H_{\rm eff}^{*\mu}
&=&
-\frac{J a_0^2}{\gamma \hbar}
({\mathcal R}^{\mu x} \tilde{s}^x +{\mathcal R}^{\mu y} \tilde{s}^y)
\nonumber
\\
&=&
-\frac{e J a_0^2}{\pi \gamma \hbar} \lambda_{{\rm so},i}^\nu
\sum_{\sigma = \pm} \bigg\{
2 m^\nu \partial_t \Big[
\Big( {\mathcal R}^{\mu x} A_{{\rm s},i}^x +{\mathcal R}^{\mu y} A_{{\rm s},i}^y \Big)
\sigma {\rm Re} C_{1,\sigma}
-\Big( {\mathcal R}^{\mu x} A_{{\rm s},i}^y -{\mathcal R}^{\mu y} A_{{\rm s},i}^x \Big)
{\rm Im} C_{1,\sigma}
\Big]
\nonumber
\\
&&
+\partial_t m^\nu \Big[
\Big( {\mathcal R}^{\mu x} A_{{\rm s},i}^x +{\mathcal R}^{\mu y} A_{{\rm s},i}^y \Big)
\sigma {\rm Re} \Big( C_{1,\sigma} +C_{2,\sigma} \Big)
-\Big( {\mathcal R}^{\mu x} A_{{\rm s},i}^y -{\mathcal R}^{\mu y} A_{{\rm s},i}^x \Big)
{\rm Im} \Big( C_{1,\sigma} -C_{2,\sigma} \Big)
\Big]
\nonumber
\\
&&
-\Big( \partial_t A_{{\rm s},i}^z +\partial_i A_{{\rm s},t}^z \Big) \Big[
\Big( {\mathcal R}^{\mu x} {\mathcal R}^{\nu x} +{\mathcal R}^{\mu y} {\mathcal R}^{\nu y} \Big)
\sigma {\rm Re} \Big( C_{1,\sigma} -C_{2,\sigma} \Big)
-\Big( {\mathcal R}^{\mu x} {\mathcal R}^{\nu y} -{\mathcal R}^{\mu y} {\mathcal R}^{\nu x} \Big)
{\rm Im} \Big( C_{1,\sigma} +C_{2,\sigma} \Big)
\Big]
\nonumber
\\
&&
+\frac{4 e}{\hbar} m^\mu m^\nu \Big[
\Big( A_{{\rm s},i}^x A_{{\rm s},t}^y -A_{{\rm s},i}^y A_{{\rm s},t}^x \Big) \sigma {\rm Re} C_{1,\sigma}
-\Big( A_{{\rm s},i}^x A_{{\rm s},t}^x +A_{{\rm s},i}^y A_{{\rm s},t}^y \Big) {\rm Im} C_{1,\sigma}
\Big]
\nonumber
\\
&&
+\frac{4 e}{\hbar} m^\nu A_{{\rm s},t}^z \Big[
\Big( {\mathcal R}^{\mu x} A_{{\rm s},i}^y -{\mathcal R}^{\mu y} A_{{\rm s},i}^x \Big) \sigma {\rm Re} C_{1,\sigma}
+\Big( {\mathcal R}^{\mu x} A_{{\rm s},i}^x +{\mathcal R}^{\mu y} A_{{\rm s},i}^y \Big) {\rm Im} C_{1,\sigma}
\Big]
\nonumber
\\
&&
+\partial_i A_{{\rm s},t}^z \Big[
\Big( {\mathcal R}^{\mu x} {\mathcal R}^{\nu x} +{\mathcal R}^{\mu y} {\mathcal R}^{\nu y} \Big)
\sigma {\rm Re} \Big( C_{1,\sigma} -C_{2,\sigma} \Big)
-\Big( {\mathcal R}^{\mu x} {\mathcal R}^{\nu y} -{\mathcal R}^{\mu y} {\mathcal R}^{\nu x} \Big)
{\rm Im} \Big( C_{1,\sigma} +C_{2,\sigma} \Big)
\Big]
\bigg\}.
\end{eqnarray}
Here we used the relations,
\begin{align}
\partial_t {\mathcal R}^{\mu x}
&=
-\frac{2 e}{\hbar} \Big( A_{{\rm s},t}^y m^\mu -A_{{\rm s},t}^z {\mathcal R}^{\mu y} \Big),
\\
\partial_i {\mathcal R}^{\mu x}
&=
\frac{2 e}{\hbar} \Big( A_{{\rm s},i}^y m^\mu -A_{{\rm s},i}^z {\mathcal R}^{\mu y} \Big),
\\
\partial_t {\mathcal R}^{\mu y}
&=
\frac{2 e}{\hbar} \Big( A_{{\rm s},t}^x m^\mu -A_{{\rm s},t}^z {\mathcal R}^{\mu x} \Big),
\\
\partial_i {\mathcal R}^{\mu y}
&=
-\frac{2 e}{\hbar} \Big( A_{{\rm s},i}^x m^\mu -A_{{\rm s},i}^z {\mathcal R}^{\mu x} \Big).
\end{align}
To make our calculation tractable and simplify notation, we define constants ${\mathcal C}_i$ as 
\begin{align}
{\mathcal C}_1
&=
-\frac{J a_0^2}{2 \pi }
{\rm Re} \sum_{\sigma = \pm} \sigma \Big( C_{1,\s} +C_{2,\s} \Big) = -\frac{Ja_0^2}{2\pi} \frac{m}{\hbar^2}\frac{\varepsilon_{\rm F}J(J^2 - \eta^2)}{2\eta(J^2 +\eta^2)^2},
\\
{\mathcal C}_2
&=
-\frac{J a_0^2}{2 \pi }
{\rm Im} \sum_{\sigma = \pm} \Big( C_{1,\s} -C_{2,\s} \Big) = \frac{Ja_0^2}{2\pi} \frac{m}{\hbar^2}\frac{ \varepsilon_{\rm F}J^2}{(J^2 +\eta^2)^2},
\\
{\mathcal C}_3
&=
-\frac{J a_0^2}{2 \pi }
{\rm Re} \sum_{\sigma = \pm} \sigma \Big( C_{1,\s} -C_{2,\s} \Big)= -\frac{Ja_0^2}{2\pi} \frac{m}{\hbar^2}\frac{\varepsilon_{\rm F}J(J^2 + 3\eta^2)}{2\eta(J^2 +\eta^2)^2} ,
\\
{\mathcal C}_4
&=
-\frac{J a_0^2}{2 \pi }
{\rm Im} \sum_{\sigma = \pm} \Big( C_{1,\s} +C_{2,\s} \Big)=  -\frac{Ja_0^2}{2\pi} \frac{m}{\hbar^2}\frac{ \varepsilon_{\rm F}\eta^2}{(J^2 +\eta^2)^2},
\\
{\mathcal C}_5
&=
-\frac{J a_0^2}{2 \pi }
{\rm Re} \sum_{\sigma = \pm} \sigma C_{1,\s} =  -\frac{Ja_0^2}{2\pi} \frac{m}{\hbar^2}\frac{ \varepsilon_{\rm F}\eta J}{(J^2 +\eta^2)^2},
\\
{\mathcal C}_6
&=
-\frac{J a_0^2}{2 \pi }
{\rm Im} \sum_{\sigma = \pm} C_{1,\s} =  -\frac{Ja_0^2}{2\pi} \frac{m}{\hbar^2}\frac{ \varepsilon_{\rm F}(J^2- \eta^2)}{2(J^2 +\eta^2)^2}.
\end{align}
From which we obtain 
\begin{eqnarray}
H_{\rm eff}^{*\mu}
&=&
-\frac{\lambda_{{\rm so},i}^\nu}{\gamma}
\bigg\{
-2 m^\nu \partial_t \Big[
{\mathcal C}_5 ({\bf m} \times \partial_i {\bf m})^\mu
-{\mathcal C}_6 \partial_i m^\mu
\Big]
+\partial_t m^\nu \Big[
{\mathcal C}_1 ({\bf m} \times \partial_i {\bf m})^\mu
-{\mathcal C}_2 \partial_i m^\mu
\Big]
\nonumber
\\
&&
+\partial_i {\bf m} \cdot ({\bf m} \times \partial_t {\bf m}) \Big[
{\mathcal C}_3 (\delta^{\mu \nu} -m^\mu m^\nu)
-{\mathcal C}_4 \epsilon^{\mu \nu o} m^o
\Big]
\nonumber
\\
&&
+2 m^\mu m^\nu \Big[
{\mathcal C}_5 \partial_i {\bf m} \cdot ({\bf m} \times \partial_t {\bf m})
+{\mathcal C}_6 \partial_i {\bf m} \cdot \partial_t {\bf m}
\Big]
\nonumber
\\
&&
-\frac{4 e}{\hbar}  m^\nu A_{{\rm s},t}^z \Big[
{\mathcal C}_5 \partial_i m^\mu
+{\mathcal C}_6 ({\bf m} \times \partial_i {\bf m})^\mu
\Big]
-\frac{2 e}{\hbar} \partial_i A_{{\rm s},t}^z \Big[
{\mathcal C}_3 (\delta^{\mu \nu} -m^\mu m^\nu)
-{\mathcal C}_4 \epsilon^{\mu \nu o} m^o
\Big]
\bigg\}.
\end{eqnarray}
(Note: the last two terms proportional to $A_{{\rm s},t}^z$ and $\partial_i A_{{\rm s},t}^z$ are expected to cancel out for gauge invariance with the other contributions we don't consider here.)
In the above calculation, we used the following relations for spin gauge field $A_{\rm s}$:
\begin{align}
{\mathcal R}^{\mu x} A_{{\rm s},i}^x +{\mathcal R}^{\mu y} A_{{\rm s},i}^y
&=
-\frac{\hbar}{2 e} ({\bf m} \times \partial_i {\bf m})^\mu,
\\
{\mathcal R}^{\mu x} A_{{\rm s},i}^y -{\mathcal R}^{\mu y} A_{{\rm s},i}^x
&=
-\frac{\hbar}{2 e} \partial_i m^\mu,
\\
\partial_t A_{{\rm s},i}^z +\partial_i A_{{\rm s},t}^z
&=
\frac{\hbar}{2 e} \partial_i {\bf m} \cdot ({\bf m} \times \partial_t {\bf m}),
\\
{\mathcal R}^{\mu x} {\mathcal R}^{\nu x} +{\mathcal R}^{\mu y} {\mathcal R}^{\nu y}
&=
\delta^{\mu \nu} -m^\mu m^\nu,
\\
{\mathcal R}^{\mu x} {\mathcal R}^{\nu y} -{\mathcal R}^{\mu y} {\mathcal R}^{\nu x}
&=
\epsilon^{\mu \nu o} m^o,
\\
A_{{\rm s},i}^x A_{{\rm s},t}^y -A_{{\rm s},i}^y A_{{\rm s},t}^x
&=
\left(\frac{\hbar}{2 e}\right)^2 \partial_i {\bf m} \cdot ({\bf m} \times \partial_t {\bf m}),
\\
A_{{\rm s},i}^x A_{{\rm s},t}^x +A_{{\rm s},i}^y A_{{\rm s},t}^y
&=
-\left(\frac{\hbar}{2 e}\right)^2 \partial_i {\bf m} \cdot \partial_t {\bf m}.
\end{align}
The spin torque is then computed using 
\begin{equation}
{\bf T} = -\gamma {\bf m}\times {\bf H}_{\rm eff}^*.
\end{equation}

\end{document}